# Theoretical Bounds on Mate-Pair Information for Accurate Genome Assembly

Henry Lin
JP Sulzberger Columbia Genome Center, Columbia University Medical Center
hcl2112@columbia.edu

**Abstract**

Over the past two decades, a series of works have aimed at studying the problem of genome assembly: the process of reconstructing a genome from sequence reads. An early formulation of the genome assembly problem showed that genome reconstruction is NP-hard when framed as finding the shortest sequence that contains all observed reads. Although this original formulation is very simplistic and does not allow for mate-pair information, subsequent formulations have also proven to be NP-hard, and/or may not be guaranteed to return a correct assembly.

In this paper, we provide an alternate perspective on the genome assembly problem by showing genome assembly is easy when provided with sufficient mate-pair information. Moreover, we quantify the number of mate-pair libraries necessary and sufficient for accurate genome assembly, in terms of the length of the longest repetitive region within a genome. In our analysis, we consider an idealized sequencing model where each mate-pair library generates pairs of error free reads with a fixed and known insert size at each position in the genome.

Even in this idealized model, we show that accurate genome reconstruction cannot be guaranteed in the worst case unless at least roughly $R/2L$ mate-pair libraries are produced, where $R$ is the length of the longest repetitive region in the genome and $L$ is the length of each read. On the other hand, if $\lceil R/L \rceil+1$ mate-pair libraries are provided, then a simple algorithm can be used to find a correct genome assembly easily in polynomial time. Although $\lceil R/L \rceil+1$ mate-pair libraries can be too much to produce in practice, the previous bounds only hold in the worst case. In our last result, we show that if additional conditions hold on a genome, a correct assembly can be guaranteed with only $O(\log(R/L))$ mate-pair libraries.

**Keywords**
Genome Assembly, De Bruijn Graph Assembly, Mate-Pair Sequencing, Paired-End Sequencing, Algorithms and Theory

# Introduction

The development of new sequencing technologies has enabled a vast amount of new genomic information to be generated at dramatically lower costs. These technologies produce short 'reads' from fragments of the genome, which later need to be assembled computationally in order to reconstruct the original genome. The main challenge of genome assembly occurs due to repeated sequences in the genome longer than the read length. To resolve the placement of repeated sequences in the genome, mate-pair sequencing is often used to sequence pairs of reads at some approximate distance apart, known as the insert size. In practice, multiple mate-pair libraries with different insert sizes are often generated to help resolve repetitive segments of different lengths within the genome.

Given the cost and effort needed to construct mate-pair libraries of different insert sizes, a natural question arises as to how many mate-pair libraries and of what insert sizes are necessary and sufficient to guarantee a correct reconstruction of the original genome. Despite the wide use of mate-pair sequencing technology over the last decade, this question has not yet been addressed from a precise theoretical perspective. We provide here the first known bounds on the amount of mate-pair information necessary and sufficient to reconstruct the genome.

Given sequencing reads of length $L$ and a genome whose longest repetitive region is of length $R$, we can construct an example where unfortunately any gap of length greater than $2L + 4$ between the insert sizes chosen can produce ambiguities in the genome assembly problem which cannot be resolved with certainty. This implies that at least roughly at least $R/2L$ insert sizes are necessary to guarantee a correct assembly of the genome (or $\lfloor R/(2L + 4) \rfloor - 1$ to be exact). On the other hand, we can provide a simple algorithm that can guarantee the correct assembly of the genome, when mate-pair information from $\lceil R/L \rceil + 1$ insert sizes is given. These two results yield the first upper and lower bounds on the mate-pair information necessary and sufficient to guarantee a correct reconstruction of the original genome.

Although producing $\lceil R/L \rceil + 1$ mate-pair libraries can be too prohibitively expensive and time consuming to produce in practice, we derive an additional condition on the genome under which a correct assembly can be guaranteed with only $O(\log (R/L))$ mate-pair libraries. To reconstruct the genome with logarithmically fewer insert sizes, we utilize a practical strategy of producing libraries with doubling insert sizes (of size $L$, $2L$, $4L$, etc.). These libraries can provably yield a correct assembly if an additional condition holds on the genome being assembled.

Although the condition is not easy to describe concisely, a simple version of the condition roughly states that reconstruction is possible with doubling mate-pair libraries when each repetitive region is adjacent to a unique region, which is larger than it (plus one read length). This simple condition does not precisely characterize when genome assembly is possible with doubling libraries, since it is easy to construct genomes which do not satisfy this condition, yet can nonetheless be assembled with doubling mate-pair libraries.

A more complicated condition described later in this paper covers a more complete set of genomes for which assembly is possible with doubling mate-pair libraries. The condition described is sufficient to guarantee a correct assembly with doubling mate-pair libraries, but may not be *necessary* for genome assembly. An open question remains if necessary and sufficient conditions can be derived, which more accurately characterize when genome assembly is possible with the doubling mate-pair strategy.

**Related Work**

The problem of genome assembly has been studied extensively for over two decades. We provide a brief summary of some related work here, but for a more complete summary, we direct readers to two recent survey papers on genome assembly [1, 2]. The first theoretical formulation of the genome reconstruction problem defined the problem as finding the shortest sequence containing all observed reads. This formulation of genome assembly is NP-hard by a reduction to the shortest superstring problem [3]. However, this model of genome assembly was too simple to accurately represent the problem of genome reconstruction. In particular, Kececioglu [4] noted the shortest sequence containing all reads may not yield the original genome, since the true genome can have the same sequence repeated many times. As a result, the shortest sequence representing the reads may not be the true genome.

This led Myers [5] to define a probabilistic model for sequencing and formulate the genome assembly problem as one of finding the sequence that is most likely to explain the reads, which was utilized in subsequent papers [6–10]. This formulation also leads the problem to be NP-hard, and unfortunately, the correct assembly may still be ambiguous, as multiple solutions may yield the same optimal likelihood score. Alternate formulations of the assembly problem utilizing the concepts of overlap graphs [4, 11], string graphs [12], or de Bruijn graphs [13] can also be shown to be NP-hard [14, 15], or may not always yield the correct assembly.

Despite the difficult nature of genome assembly, a number of genome assemblers (such as the early Celera assembler [16] or more recent assemblers [17–23]) have been developed to work in practice; see [1, 2] for a more complete survey of genome assemblers. Modern assemblers require the use of heuristics in order to overcome the complexity challenges of genome assembly, but unfortunately may cause assembly errors, as recent genome assembly evaluations have shown [24–26]. For these reasons, we seek to study the potential for genome assemblers to produce provably correct assemblies, when provided with varying amounts of mate-pair information.

There has been work showing that formulations of the assembly problem with mate-pair information are also NP-hard [27, 28], including recent work [29] characterizing the parametric complexity of the problem in terms of various parameters such as read lengths and the length of repetitive regions. Additionally, Wetzel et al. [30] sought to study the benefits of tuning the choice of mate-pair sizes to optimize the assembly of a genome. Their work is primarily empirical in nature, evaluating the benefits of tuning mate-pair insert sizes in a simulated setting. Our work proceeds from a more theoretical perspective, seeking to bound the minimum number of insert sizes needed to guarantee correct reconstruction of a genome.

Our work has a similar goal to the recent work by Motahari et al. [31] and Bresler et al. [32] which characterizes the amount of sequencing information required for accurate genome assembly with single-end reads. Our work differs from theirs by focusing on sequencing with mate-pair libraries. Previously, Ukkonen [33] also showed conditions necessary and sufficient for genome assembly when single-end reads are used, but the results do not consider the use of mate-pair read libraries.

Note that sequencing with mate-pairs creates subtle differences from the single-end read case, which must be accounted for in the proofs. As one example, sequencing mate-pair reads of length L and insert sizes of length 0, L, 2L, ... , rL is not equivalent to having single-end reads of the longest fragment length (rL+2L). The latter case with longer reads actually provides more information (assuming uniform coverage over the genome), and one can construct an example where reconstruction is possible in the latter case, but not in the former case.

**Model**

In this paper, we model the problem of sequence assembly from a theoretical perspective without modeling additional challenges that may occur in practice, such as errors in the reads. Our model simplifications mean that our theoretical formulation of the sequence assembly problem is an easier problem than the one that occurs in practice. However, note that our lower bound on the theoretical sequence assembly problem still has practical implications for the even more challenging task of sequence assembly in the presence of errors and other real-world conditions.

For our theoretical model of sequence assembly, we ignore three challenges that make sequence assembly more challenging in practice: errors in the reads, uneven coverage of reads, and uncertainty in the mate-pair insert size. As sequencing technology improves, these factors may become less important, as we have seen substantial improvements in reducing the amount of error and improving the evenness of coverage when sequencing reads; additionally, estimates of the insert size of mate-pairs may also improve in the future. For those reasons, in this paper we will consider an idealized model where we assume each read does not contain errors, each mate-pair library generates pairs of reads with a fixed and known insert size, and each library produces one mate-pair starting from every location in the genome. In this idealized model, we are better able to study the theoretical limits of sequence assembly.

**Notation of Repeats and Repetitive Regions**

Before stating our main theorem, we first need to define formally some terminology related to the repetitive nature of genomes that makes genome assembly difficult. The first factor that influences the difficulty of genome assembly is the length of the longest sequence that appears more than once in the genome, which we will denote by $R_M$. It is not hard to show that if the read length is at least $R_M+2$, then genome reconstruction is easy to solve, assuming complete coverage of the genome. (When reads are longer than $R_M+2$, one can construct a de Bruijn graph with each edge representing a sequence overlap of length $R_M+1$; then since sequences of length $R_M+1$ occur uniquely in the genome, each node in the de Bruijn graph will have in-degree and out-degree at most 1, and the genome can be reconstructed easily by traversing the simple path created by the de Bruijn graph).

When the read length L is smaller than $R_M+2$, then the genome reconstruction problem is more difficult and requires mate-pair information. In this setting, the difficulty of the genome reconstruction problem depends more on the length of the longest *repetitive region*, defined below, rather than $R_M$, the length of the longest *repeated sequence* in the genome.

For a read length L, we first define a unique (or non-repetitive) region of the genome to be any consecutive sequence S such that all substrings of length L-1 occur uniquely in the genome. We define unique regions using substrings of length L-1, since unique regions are easy to reconstruct using the de Bruijn graph approach with reads of length L. Next, we define any consecutive sequence in the genome which does not overlap or contain a unique region, to be a repetitive region (regions in which all substrings of S of length L-1 appear at least twice in the genome). Note that the unique and repetitive regions in the genome must be defined with respect to a particular read length L, as different read lengths will produce different unique and repetitive regions. Given these definitions, we will define R to be the length of the longest repetitive region occurring in the genome with respect to read length L, and quantify the difficulty of genome assembly in terms of R and L.

In the next two sections, we provide the proofs of our two main results for reconstruction of a genome with length L reads and maximum repetitive region length R. The first result shows that $\lfloor R/(2L + 4) \rfloor - 1$ mate-pair libraries are necessary to guarantee a correct assembly. The second result shows that $\lceil R/L \rceil + 1$ mate-pair libraries are sufficient to guarantee a correct assembly.

**A lower bound on mate-pair information necessary for genome assembly**

**Theorem 1:** Let $S_1, S_2, \ldots, S_k$ be the insert sizes of k mate-pair libraries with read length L. If there exists any gap between insert sizes $S_i - S_{i-1}$ of length greater than 2L+4, then there exists a genome whose longest repetitive region is less $S_k$ yet the genome cannot be reconstructed. In particular, there exists two different genomes with longest repetitive region at most $S_k$, which yield identical mate-pair information for insert sizes $S_1, S_2, \ldots, S_k$.

**Corollary 1:** In order to guarantee a correct genome assembly for a genome whose longest repetitive region is size R, at least $\lfloor R/(2L + 4) \rfloor - 1$ mate-pair libraries are needed.

To show that the genome reconstruction problem cannot be solved in this case, we will define two genome sequences that yield the exact same mate-pair information in our idealized model. As a result, no algorithm can distinguish between which of the two genomes is the correct reconstruction.

The illustration below shows the construction of the two distinct sequences which yield the exact same mate-pair information for insert sizes $S_1, S_2, \ldots, S_k$. In the illustration below, the blue, orange, and red blocks represent sequences of two nucleotides, while the green blocks represent repetitive regions to be defined later. The first indistinguishable sequence $G_{CT}$ is the concatenation of the sequences in the blocks in the 2 rows shown below with the (blue) CC sequence inserted in the first row and (red) TT sequence in the second row. The second indistinguishable sequence $G_{TC}$ is the same sequence except with the (red) TT sequence inserted in the first row and the blue (CC) sequence inserted in the second row. For clarity, the last R1 block in the first row is repeated as the first block in the second row for illustration purposes, but only occurs once at that position. By defining the sequences for the green repetitive blocks appropriately, we can show that both $G_{CT}$ and $G_{TC}$ sequences yield the same mate-pair information, provided there is a gap of length greater than 2L+4 between two mate-pair libraries $S_i$ and $S_{i-1}$.

In the construction below, we define the green block labeled "R2" to be a sequence of $(S_{i-1}+2L)$ consecutive A's (a homopolymer run), while the green block labeled "R1" is a consecutive sequence of $(S_k - S_{i-1})$ G's. Note that the repetitive regions of the constructed genomes are of size at most $S_k$, a property required of the theorem.

Now to prove that both $G_{CT}$ and $G_{TC}$ sequences yield the same mate-pair information, we need to examine all mate-pair information involving the ambiguous middle blocks in the two rows, comprised of either the CC or TT sequence. Note that there is an absence of any mate-pair information from the ambiguous CC/TT blocks to the unique GC, GT, TG, CG sequences shown in orange, since those are at distance at least $(S_{i-1}+2L)$. The mate-pair libraries of size $S_{i-1}$ or smaller are too small to span repeat "R2" and the mate-pair libraries of size larger than $S_i$ or larger cannot connect the ambiguous CC/TT blocks to the neighboring orange blocks either, since the insert size $S_i$ is at least 4 base pairs larger than "R2". Also note that the largest insert size $S_k$ is too small to connect the ambiguous CC/TT blocks to any orange block on a different row, since they are at a distance of at least $S_k+2L$ as well.

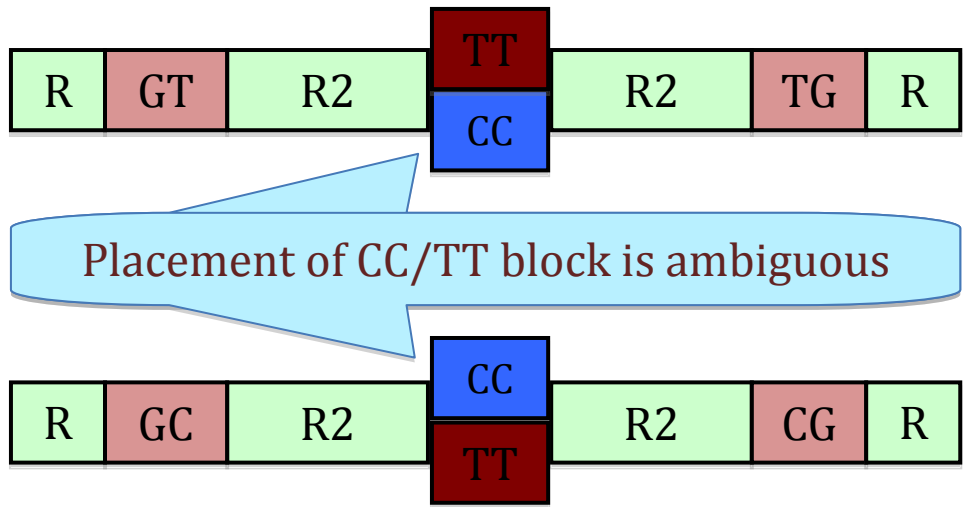

**Figure 1.** An illustration of a genome that is difficult to reconstruct when mate-pair information from the CC/TT blocks to the unique orange blocks are missing.

Thus, all mate-pair information from the ambiguous CC/TT blocks hit repeat regions comprised of homopolymer runs of A's or G's. As a result, either ordering of the ambiguous CC/TT blocks produces the exact same set of mate-pair information, and no assembly algorithm can distinguish the difference between the sequences $G_{CT}$ and $G_{TC}$ with the provided mate-pair information.

**An upper bound on mate-pair information necessary for genome assembly**

**Theorem 2:** Given reads of length L and a genome whose longest repetitive region is R (with respect to read length L), sequencing mate-pair libraries of sizes 0, L, 2L, 3L, ..., rL is sufficient to reconstruct the genome, where r = $\lceil R/L \rceil$.

**Corollary 2:** For a genome with longest repetitive region R, $\lceil R/L \rceil$+1 mate-pair libraries are sufficient for genome reconstruction.

To show that we can always reconstruct the genome when the above mate-pair libraries are given, we first derive a procedure to reconstruct the unique regions of the genome, and then apply a method to reconstruct the repetitive regions between the unique regions of the genome with mate-pair information. The strategy we employ is similar to the strategy employed by the Celera assembler [16], which constructs "unitigs" (uniquely assemble-able contigs) and reconstructs the repetitive regions between unitigs with mate-pair information. We omit the proof due to space limitations, but provide formal details in the Appendix. In the Appendix, we also provide two methods for constructing the set of unitigs (i.e. the unique regions of the genome). The first method uses coverage information and requires the assumption that reads are generated uniformly from the genome with one read sequenced at each location. However, we also note in the Appendix that a more robust procedure that can be derived which can reconstruct the unique regions in the genome without using assuming exactly uniform coverage of the genome.

**Genome Assembly with Doubling Mate-Pair Libraries**

In this section, we show that creating mate-pair libraries with doubling insert sizes is sufficient with additional conditions on genome being assembled. In our analysis, we assume that mate-pair libraries of read length L and insert sizes 0, L, 2L, 4L, …, $2^r$L are produced, where r is the smallest integer such that $2^r$L is greater than or equal to R, the length of the longest repetitive region. To fully construct our genome, we will start by constructing the unique regions of the genome by using a de Bruijn graph approach (see approach used in proof of Theorem 2 in the Appendix). Then we will utilize an additional condition on the genome, which we show can guarantee a complete genome assembly with the doubling mate-pair libraries.

Before deriving a more complicated condition under which a complete genome assembly is guaranteed, we start by showing a simple condition that is sufficient for assembly, which will illustrate the main idea behind the more general and complicated sufficient condition described later. Our simple condition that makes doubling insert sizes sufficient for genome assembly states that each repetitive region is adjacent to a unique region, which is at least L bases longer than it is. The main intuition as to why long unique regions are helpful for resolving neighboring repetitive regions is because one can show that unique regions can be used to resolve neighboring repetitive regions of roughly the same size with the given doubling mate-pair libraries.

In particular, we can show that if we have constructed a unique region of length U we can resolve neighboring repetitive regions of roughly the same size using mate-pair libraries of size $M_1 = L*2^i$ and $M_2 = L*2^{i+1}$, where i is chosen such that $M_1 ≤ (U–L) < M_2$. The exact number of bases we will show we can reconstruct before or after the unique region will be $M_2 + L$, which will allow us to resolve any neighboring repetitive regions of length $M_2$ or less (with L bases being used to determine which unique region follows the current repetitive region). Since $M_2$ is at least U–L by construction, this means that we can resolve repetitive regions of size U–L or less, or in other words a repetitive region can be resolved provided that it is adjacent to a unique region at least L bases longer than it is.

We now describe how the next $M_2 + L$ bases following the unique region of length U can be reconstructed using two mate-pair libraries defined above (and the previous $M_2 + L$ bases occurring before the unique region can also be resolved in a similar manner). At a high level, we will show that the smaller mate-pair library can be used to reconstruct the first $M_1 + L$ bases after the end of the unique region, while the larger mate-pair library can be used to reconstruct the next $M_1$ bases after that, so in total we can reconstruct $M_2 + L$ bases after the end of the unique region (as $M_1=2*M_2$).

To reconstruct the first $M_1 + L$ bases, note that if we sequence a mate-pair of insert size $M_1$, starting $M_1+L$ bases before the end of the unique region, the first mate-pair will be "anchored" within the unique region, while the second mate-pair will contain the sequence of the first L bases after the unique region. This mate-pair allows us to recover the first L bases after the unique region, and we can recover the next base after that by examining the mate-pair which appears $(M_1+L)-1$ before the end of the unique region. Furthermore, we can proceed in a similar manner until we reach the last L bases of the unique region, and have recovered $M_1+L$ bases after the end of the unique region.

Similarly, we can use a mate-pair of insert size $M_2$, starting $M_1$ bases before the end of the unique region to recover the next L bases, after the previously recovered $M_1+L$ bases. Proceeding in the same manner as before with mate-pairs of insert size $M_2$, we can thus recover in total $M_2 + L$ bases after the end of the unique region, and as a result, resolve any repetitive region of length at most $M_2$ with the two mate-pair libraries defined (where the L final bases may be used to determine which unique region follows the current one).

Note that the simple condition defined above is inefficient in the sense that a unique region can also resolve non-adjacent repetitive regions if the unique region is sufficiently long and has the requisite doubling mate-pair information. To define a more precise condition under which genome assembly is possible with doubling mate-pair libraries, we will define a graph based on the unique regions in the genome, and if the final graph is connected, we can guarantee a correct genome reconstruction with the doubling mate-pair strategy.

For each unique region in our genome, we construct a node in our graph to represent it. If the genome starts with a repetitive sequence, we also add a node to represent the start of the genome, and similarly if the genome ends on a repetitive region, we add a node for the end of the genome (with these nodes representing unique regions of size 0). Now, we connect two nodes in the graph representing unique regions of size $U_1$ and $U_2$, if the unique regions are separated by a distance that is less than $L*2^i$, where i is the smallest integer such that $(max(U_1,U_2) - L) < L*2^i$. (In case $(max(U_1,U_2) - L)$ exceeds the largest insert size $2^r L$, then we just set i=r in the previous statement). By defining edges in such a manner, we can see that nodes with edges between them can be bridged with the mate-pair strategy described above. After constructing this graph, note that if the final graph is connected then the repetitive regions between each pair of unique regions can be resolved with the mate-pair information provided, and thus the entire genome can be reconstructed.

**Open Questions**

We have shown bounds on the amount of mate-pair information necessary and sufficient for genome assembly. One immediate open question is whether or not the lower bound or upper bound can be improved, so that we have matching upper and lower bounds. It is unclear whether or not the true upper and lower bounds should be closer to R/L or R/2L, respectively.

Additionally, those bounds hold in the worst-case, and it is unclear how often these cases might occur in practice. We have shown additional conditions on the genome can make assembly possible when utilizing mate-pairs with doubling insert sizes. However, the conditions shown are only sufficient, but may not be necessary for genome reconstruction. An interesting open question would be to search for conditions that are both necessary and sufficient for genome assembly, which better characterizes when assembly is possible with double mate-pair libraries.

Furthermore it would be interesting to consider developing good adaptive strategies for determining the best insert size(s) to use for mate-pair sequencing. In particular, one might perform an initial sequencing experiment to gain insight into the repeat structure of the genome, for example by constructing a de Bruijn graph, and then only choose insert sizes for future mate-pair library construction after analyzing the structure of the initial de Bruijn graph created.

Lastly, our work considers an idealized model, which assumes that mate-pair libraries are generated with a fixed and known insert size, rather than being created from a distribution. Our result showing that any gap of length at least $2L+4$ in mate-pair insert sizes can cause the genome assembly problem to have an ambiguous solution, still applies when considering the more realistic insert-size model. When we have a distribution of insert sizes for each library, we can no longer conclude that $\lfloor R/(2L + 4) \rfloor - 1$ mate-pair libraries are needed, although we can produce a rough lower bound if we know that each library generates a distribution of mate-pair insert sizes over a fixed range of at most $D$ base pairs. With our previous result showing that gaps of more than $2L+4$ in mate-pair insert information leads to ambiguities in genome assembly, one can conclude at least $\lfloor R/(D + 2L + 4) \rfloor - 1$ mate-pair libraries are required to guarantee a correct reconstruction in this case. A more precise bound improving on the rough analysis above would also be interesting.

## Competing Interests

The author has no competing interests.

## Acknowledgement

The author would like to acknowledge Yufeng Shen and Mihai Pop for helpful discussions on genome assembly. The author also acknowledges Itsik Pe'er, Michael Schatz, and Mohammadreza Ghodsi for helpful feedback on an initial draft of the paper.

# Appendix

**Proof of Theorem 2:**

A standard way to construct unique regions in the genome (i.e. unitigs) is to construct a de Bruijn graph with nodes representing sequences of length L (observed in the reads) and edges connecting nodes with overlapping sequences of length L-1. Then any path from a node u to a node v in the graph whose intermediate nodes all have in-degree and out-degree one can be compressed into a single edge from u to v, where the edge represents the composite sequence of all the nodes in the compressed path. After all nodes of in-degree and out-degree one have been compressed, the edges representing the composite sequences of compressed paths are represent contiguous sequences that occur in the genome (otherwise known as contigs). These contigs may occur one or more times in the genome, but if uniform coverage is assumed, the sequences coming from the unique regions in the genome) can be distinguished by the number of reads covering the contig. If we assume that exactly one read is sequenced from each location in the genome, then we can determine the contigs representing sequences from unique regions in genome by only including the contigs with only one read covering each sequence of length L within the contig.

Alternatively, there is also another method that can reconstruct the unique regions using the mate-pair information provided by Theorem 2. This second method does not require exactly even coverage of the genome, and can handle the case where more than one read may be sequenced from each location. To reconstruct the unique regions in this case, we start by determining for each read with sequence s if sequence s occurs uniquely in the genome. To do so, we consider all mate-pairs, whose first mate-pair read has sequence s. We consider the sequence s to have a mate-pair conflict, if it appears in two mate-pairs with the same insert size, yet their second mate-pair consists of two different sequences. Note that if a sequence s has a mate-pair conflict then it must come from a repetitive region, since our model sequences reads without errors in the read or insert size.

By determining the set of sequences that do not have any mate-pair conflict, we can thus determine the S set of all sequences of length L that are unique in the genome. With the sequences in set S, we can then construct a de Bruijn graph of order L (nodes and edges representing sequences of length L and L-1 respectively). Although the sequences in S occur uniquely in the genome, some nodes in the de Bruijn graph may still have in-degree or out-degree greater than 1, since edges represent overlapping sequences of length L-1 and may not be unique. However, if we delete the outgoing edges of nodes with out-degree greater than one and delete the incoming edges of nodes with in-degree greater than one, then the remaining edges will represent unique sequence overlaps of length L-1. Furthermore, the nodes in the remaining de Bruijn graph will have in-degree and out-degree at most 1, so that the graph will consists only of simple (disjoint) paths. These paths can then be traversed in order to find all sequences from unique regions in the genome.

Once the unique regions from the genome have been reconstructed, we can then reconstruct the repetitive regions between unique regions with mate-pair information. For each unique segment, we extract all mate-pair reads, whose first mate-pair read has the same sequence s as the last L characters of the unique segment. The mate-pairs associated with the sequence s can then be used to reconstruct the next rL characters in the genome from the mate-pair reads with insert sizes 0, L, 2L, 3L, ..., rL.

Since the maximum repetitive region is size at most rL, we must have at least one mate-pair whose second read consists of a (unique) sequence appearing in the next unique region following the unique region containing the first read. The mate-pair with the smallest insert size whose secondary read appears in another contiguous unique region can be used to determine the unique region which follows the current unique region, and the mate-pairs with smaller insert sizes can reconstruct the repetitive region in between those two unique regions. By repeating this procedure for all unique segments, we can then connect all unique segments while resolving the intervening repetitive regions. Repetitive regions at the start or end of the genome can also be resolved in a similar manner. Upon resolving all repetitive regions, we have then reconstructed the entire genome.